\documentstyle[aps,twocolumn,epsf]{revtex}
\begin{document}

\title{Magnetization Switching in Small Ferromagnetic Particles:\\
  Nucleation and Coherent Rotation}
\author{U.~Nowak and D.~Hinzke}
\address{Theoretische Tieftemperaturphysik,
  Gerhard-Mercator-Universit\"{a}t-Duisburg, 47048 Duisburg/ Germany\\ 
  e-mail: uli@thp.uni-duisburg.de }

\date{August 11, 1998}
\maketitle

\begin{abstract}
  The mechanisms of thermally activated magnetization switching in
  small ferromagnetic particles driven by an external magnetic field
  are investigated. For low uniaxial anisotropy the spins rotate
  coherently while for sufficiently large uniaxial anisotropy they
  behave Ising-like, i.~e.~the switching then is due to nucleation.
  The crossover from coherent rotation to nucleation is studied for
  the classical three-dimensional Heisenberg model with uniaxial
  anisotropy by Monte Carlo simulations.  From the temperature
  dependence of the metastable lifetime the energy barrier of a
  switching process can be determined.  For the case of infinite
  anisotropy we compare numerical results from simulations of the
  Ising model with theoretical results for energy barriers for both,
  single-droplet and multi-droplet nucleation.  The simulated barriers
  are in agreement with the theoretical predictions.
\end{abstract}


With decreasing size ferromagnetic particles become single-domain
which improves their quality for magnetic recording. On the other
hand, when the particles are too small they become superparamagnetic
and then due to thermal fluctuations no magnetic information can be
stored (see e.~g.~\cite{chantrell} for a review). Hence, much effort
has been devoted to an understanding of the behavior of small magnetic
particles experimentally \cite{salling,lederman,wernsdorfer}, analytically
\cite{coffey}, and in computer simulations. In the latter case, mainly
magnetization switching by nucleation \cite{rikvold,richards,hinzke}
has been studied using Ising models of finite size, but also other
reversal mechanisms in models with continuous degrees of freedom like coherent
rotation, single-droplet nucleation, and multi-droplet nucleation have
been discussed \cite{gonzales,hinzke}.

In this paper we focus on different thermally activated reversal
mechanisms occuring in ferromagnetic particles.
We will consider a finite, spherical,
three-dimensional system of magnetic moments.  These magnetic moments
may represent atomic spins or also block spins following from a coarse
graining of the physical lattice \cite{nowak}.  Our system is defined
by a classical Heisenberg Hamiltonian,
\begin{equation}
  {\cal H} = - J \sum_{\langle ij \rangle} {\bf S}_i \cdot {\bf S}_j
  -d \sum_i (S^z_i)^2 -{\bf B} \cdot \sum_i {\bf S}_i,
\label{e:ham}
\end{equation}
where the ${\bf S}_i$ are three dimensional vectors of unit length.
The first sum which represents the exchange of the spins is over
nearest neighbors with the exchange coupling constant $J$. The second
sum represents an uniaxial anisotropy which favors the $z$-axis as easy
axis (anisotropy constant $d>0$). The last sum is the
coupling of the spins to an applied magnetic field, where ${\bf B}$ is
the strength of the field times the absolute value of the magnetic
moment of the spin. We neglect the dipolar interaction.  Therefore,
the validity of our results is restricted to particles which are small
enough to be single-domain in the remanent state \cite{aharony}.

In the following we will investigate the thermally activated reversal
of the magnetization of a particle which is destabilized by a magnetic
field pointing into the direction antiparallel to the initial
magnetization which is parallel to the easy axis of the system
initially. In this case after some time the particle will reverse its
magnetization.

Due to the many degrees of freedom of a spin system numerical methods
have to be used for a detailed microscopic description.
Since we are especially interested in the thermal properties we use
Monte Carlo(MC) methods for the simulation. Although a direct mapping
of the time scale of a MC simulation on experimental time scales is
not possible this method provides information on the dynamical
behavior since it solves the master equation for the irreversible
behavior of the system.

\begin{figure}
  \begin{center}
    \hspace*{0.5cm}
    \epsfxsize=5.5cm
    \epsffile{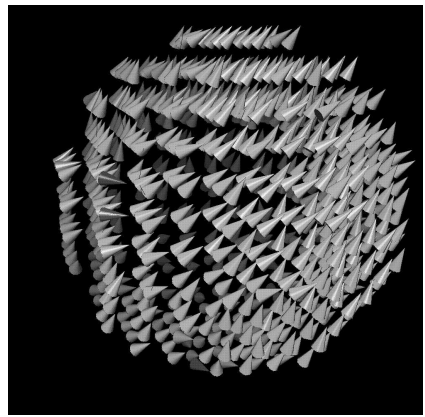}
  \end{center}
  \caption{Snap-shot of a simulated spin system during 
    coherent rotation. In the picture we omit the interior part of the
    system and show only the outer spins of the sphere. 
    $R=6$ spins, $B = 0.7J$, $d = 0.35J$/spin, $T=0.09J$. }
  \label{f:spin_cr}
\end{figure}

We consider spins on a simple cubic lattice of size $L\times L \times
L$ and simulate spherical particles with radius $R = L/2$ and open
boundary conditions on this lattice using the MC algorithm
described in \cite{hinzke}.  We start our simulations with an initial
spin configuration where all spins are pointing up ( ${\bf
  S}_i = (0,0,1)$). The magnetic field ${\bf B} = (0,0,-B)$
destabilizes the system and after some time the magnetization 
will reverse. The metastable lifetime $\tau$ is defined by the
condition $M_z(\tau) = 0$ where $M_z$ is the $z$-component of the
magnetization ${\bf M} = (1/N) \sum_i {\bf S}_i$.

For sufficient low anisotropy the spins can be expected to rotate
coherently. Such a reversal process is shown in Fig.\ref{f:spin_cr}
where a spin configuration of a simulated system of size $R=6$ spins
during the reversal process is represented. The spins are nearly
parallel, except of thermal fluctuations.  Following the
theory of N\'{e}el \cite{neel} and Brown \cite{brown} the energy
barrier which has to be overcome during the reversal is only due to
the anisotropy of the system,
\begin{equation}
  \Delta E_{cr}  =  \frac{4 \pi R^3d}{3}-{\frac {4\pi R^3BM}{3}}
  +\frac{\pi R^3B^2M^2}{3d},
  \label{e:ecr}
\end{equation}
where $M$ is the saturation magnetization of the system. 

\begin{figure}
  \begin{center}
    \hspace*{0.5cm}
    \epsfxsize=5.5cm
    \epsffile{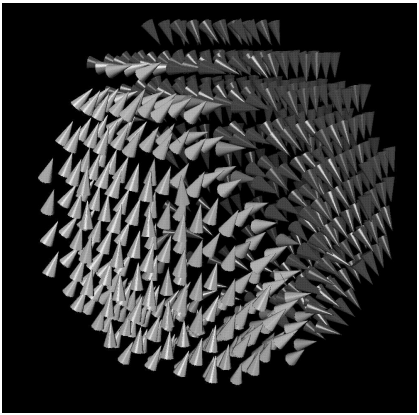}
  \end{center}
  \caption{Snap-shot of a simulated spin system as in
    Fig. \ref{f:spin_cr} but for single-droplet nucleation. The
    $z-$component of the spin is color-coded (lighter grey corresponds
    to spins-up, dark to spin-down). $B = 0.7J$, $d = 0.7J$/spin and
    $T = 0.45J$.}
  \label{f:spin_nu}
\end{figure}

For a system with a sufficient large anisotropy it is energetically
favorable to divide into parts with opposite directions of
magnetization parallel to the easy axis in order to minimize the
anisotropy energy barrier. This kind of reversal mechanism is called
nucleation \cite{becker} (see \cite{rikvold} for a recent review). The
simplest case of a reversal process driven by nucleation for a system
of finite size is the growth of one single droplet starting somewhere
at the boundary. Due to the growth of the droplet a
domain wall will cross the system and the energy barrier which has to
be overcome is caused by the domain wall energy. This is shown in
Fig.~\ref{f:spin_nu}.  Here the domain wall is in the center of the
system dividing the particle into two oppositely magnetized parts of
equal size. 

The energy barrier $\Delta E_n$ which has to be overcome during the
reversal by nucleation of one single droplet in a three dimensional
system with open boundary condition has been derived in \cite{hinzke}.
It can be expanded with respect to $MBR/\sigma$ resulting in
\begin{equation}
  \Delta E_n \approx 2 \pi R^2 \sigma - \frac{4 \pi B R^3 M}3 
  +\frac {3\pi B^2R^4 M^2}{8\sigma} + \ldots
  \label{e:enu}
\end{equation}
where $\sigma$ is the domain wall energy density.  The corresponding
lifetime of the metastable state for the single-droplet nucleation is
then
\begin{equation}
  \tau \sim \exp \left( \frac{\Delta E_{n}}{T} \right)
  \label{e:activation_nu}
\end{equation}
for temperatures $T \ll \Delta E_{n}$. Eq.~\ref{e:enu} has an
interesting interpretation: for small fields $B$ the energy barrier of a
nucleation process is the energy of a flat domain wall in the center
of the particle plus corrections which start linearly in $B$.  In
contrast to the N\'{e}el-Brown theory, here for vanishing magnetic
field the energy barrier is proportional to the cross-sectional area
of the particle rather than to its volume.

Hence, comparing the two energy barriers for coherent rotation
(Eq.~\ref{e:ecr}) and nucleation (Eq.~\ref{e:enu}) one can evaluate
the critical particle size $R_c$
where the crossover from coherent rotation to nucleation sets in.
For vanishing magnetic field this critical particle size is given by
\begin{equation}
  R_c = \frac{3\sigma}{2d}.
  \label{e:dclb}
\end{equation}
For particles larger than $R_c$ reversal by nucleation has the lower
energy barrier while the opposite is true for particles smaller than $R_c$.

Another possibility of a reversal process driven by nucleation is the growth
of several droplets at the same time (multi-droplet
nucleation). This reversal process occurs when the probability for the
growth of large droplets is high enough and when the critical droplet
size is smaller than the system size so that many droplets may occur
in the system at the same time. This is the case for higher fields or larger
temperatures \cite{rikvold2,acharyya} . The lifetime for the
multi-droplet nucleation in $D$-dimensions is \cite{becker}
\begin{equation}
  \tau \sim \exp \left( \frac{\Delta E_{n}}{(D+1)T} \right).
  \label{e:activation_md}
\end{equation}

Hence, the exponent of the exponential function changes by a factor of
$1/(D+1)$ which is 1/4 in our case. In order to quantify the crossover
from single droplet to multi-droplet nucleation we consider the case
of infinite anisotropy in which case we can directly simulate an Ising
system. We performed a standard MC simulation of the Ising 
Hamiltonian
\begin{equation}
  {\cal H} = - J \sum_{\langle ij \rangle} S_i S_j - B \sum_i S_i
\end{equation}
with $S_i = \pm 1$.

\begin{figure}
  \begin{center}
    \hspace*{0.5cm}
    \epsfxsize=5.5cm
    \epsffile{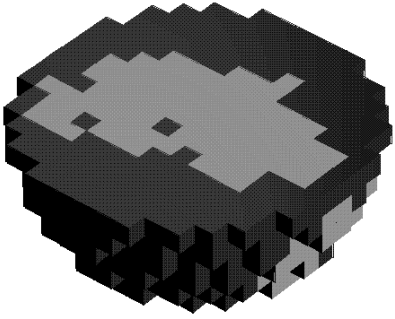}
  \end{center}
  \caption{Snap-shot of a simulated spin system at the lifetime
    $\tau$ for multi-droplet nucleation.  Shown is an Ising system of
    size $R=8$ spins. In order to show the interior of the system we
    have cut the sphere and show only one half of the system. $B =
    0.5J$ and $T = 2.8J$.}
  \label{f:spin_md}
\end{figure}

Fig.~\ref{f:spin_md} shows a corresponding spin configuration 
during the reversal. The Ising spins are represented as light (spin
up) and dark (spin down) grey boxes. The reversal process is initiated
by many droplets at the boundary of the system. Later all these
droplets join each other so that the outer shell of the particle is
reversed while the inner parts have still the initial magnetization
direction.

\begin{figure}
  \begin{center}
    \epsfxsize=7cm
    \epsffile{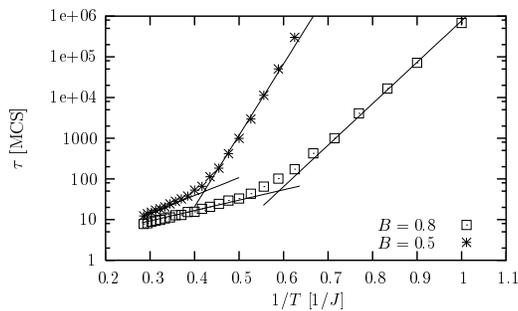}
  \end{center}
  \caption{Metastable lifetime $\tau$ vs. $1/T$ for two different 
   magnetic fields. System size $R=4$ spins.}
    \label{f:tau_ising_md}
\end{figure}

Fig.~\ref{f:tau_ising_md} shows the temperature dependence of the
metastable lifetime following from simulations for two different
fields. Each data point is an average over 100 independent runs. 
The data show a linear behavior with different slopes for low and high
temperatures, respectively. For low temperatures single-droplet
nucleation occurs and we can compare the corresponding data with the
theoretical prediction, Eqs.~(\ref{e:enu}, \ref{e:activation_nu}). The
slopes of the straight lines shown are the energy barriers $\Delta
E_n$ obtained from Eq.~\ref{e:enu} which have the theoretical values 
$40.4J$ ($B = 0.5J$) and $23.5 J$ ($B = 0.8 J$) for the system size
used here ($R = 4$ spins) and for $\sigma = 1.2J$/bond \cite{hinzke}
showing very good aggreement between the numerical data and the
theoretical predictions. For higher temperatures we find a crossover
to multi-droplet nucleation in which case theory predicts a reduction
of the energy barrier by a factor of $(D+1) = 4$,
Eq.~\ref{e:activation_md}. The slopes of the straight lines shown are
the energy barriers $\Delta E_n/(D+1)$ obtained from theory showing
again very good agreement with the simulated data.
Note that the position of all lines are fitted since we do not know
the prefactors in Eqs.~\ref{e:activation_nu} and
\ref{e:activation_md}.

To conclude we found very
good agreement between the energy barriers obtained from the numerical
data and the predictions obtained from both single- and multi-droplet
nucleation theory. Especially, the general result that the
effective energy barriers for single- and multi-droplet nucleation are
related by the $D-$dependent factor is confirmed. A
corresponding behavior concerning the field dependence of the the
metastable lifetime was also observed recently \cite{rikvold2,acharyya}.

{\bf Acknowledgments:}
We thank M.~Acharyya and K.~D.~Usadel for
helpful discussions. The work was supported by the Deutsche
Forschungsgemeinschaft through Sonderforschungsbereich 166 and through
the Graduiertenkolleg "Struktur und Dynamik heterogener Systeme".


\begin{references}
\bibitem{chantrell} R.~W.~Chantrell and K.~O'Grady, in  \emph{Applied
  Magnetism}, edited by R.~Gerber, C.~D.~Wright, and G.~Asti (Kluwer
  Academic Publishers, Dordrecht, 1994), p. 113
\bibitem{salling} C.~Salling, R.~O'Barr, S.~Schultz, I.~McFadyen, and
  M.~Ozaki. J.~Appl.~Phys. {\bf 75}, 7986 (1994)
\bibitem{lederman} M.~Lederman, S.~Schultz, and M.~Ozaki,
  Phys.~Rev.~Lett. {\bf 73}, 1986 (1994) 
\bibitem{wernsdorfer} W.~Wernsdorfer, K.~Hasselbach, D.~Mailly,
  B.~Barbara, A.~Benoit, L.~Thomas, and G.~Suran,
  J.~Mag.~Mag.~Mat. {\bf 140}, 389 (1995); W.~Wernsdorfer, E.~Bonet
  Orozco, K.~Hasselbach, A.~Benoit, B.~Barbara, N.~Demoncy,
  A.~Loiseau, H.~Pascard, and D.~Mailly, Phys.~Rev.~Lett. {\bf 78},
  1791 (1997) 
\bibitem{coffey} W.~T.~Coffey, D.~S.~F.~Crothers, J.~L.~Dormann,
  Yu.~P.~Kalmykov, E.~C.~Kennedey, and W.~Wernsdorfer,
  Phys.~Rev.~Lett., in press
\bibitem{rikvold} P.~A.~Rikvold and B.~M.~Gorman, in  \emph{Annual
    Reviews of  Computational Physics I}, edited by D.~Staufer
  (World Scientific, Singapore, 1994), p. 149
\bibitem{richards} H.~L.~Richards, M.~Kolesik, P.~A.~Lindg{\aa}rd,
  P.~A.~Rikvold, and M.~A.~Novotny, Phys.~Rev.~B {\bf 55}, 11521 (1997)
\bibitem{hinzke} D.~Hinzke and U.~Nowak, Phys.~Rev.~B {\bf 58}, 265
  (1998)
\bibitem{gonzales} J.~M.~Gonz\'{a}les, R.~Ram\'{\i}rez, R.~Smirnov-Rueda, and
  J.~Gonz\'{a}lez, Phys.~Rev.~B {\bf 52}, 16034 (1995)
\bibitem{nowak} U.~Nowak, J.~Heimel, T.~Kleinefeld, and D.~Weller,
  Phys.~Rev.~B {\bf 56}, 8143 (1997)
\bibitem{aharony} A.~Aharoni, J.~Appl.~Phys. {\bf 63}, 5879 (1988)
\bibitem{neel} L.~N\'{e}el, Ann.~Geophys. {\bf 5}, 99 (1949)
\bibitem{brown} W.~F.~Brown, Phys.~Rev. {\bf 130}, 1677 (1963)
\bibitem{becker} R.~Becker and W.~D\"{o}ring, Ann.~Phys. (Leipzig)
  {\bf 24}, 719 (1935)
\bibitem{rikvold2} P.~A.~Rikvold, H.~Tomita, S.~Miyashita, and
  S.~W.~Sides, Phys.~Rev.~E {\bf 49}, 5080 (1994) 
\bibitem{acharyya} M.~Acharyya and D.~Stauffer to be published in
  E.~Phys.~J. {\bf B}
\end{references}
\end{document}